\UseRawInputEncoding
\documentclass[aip,graphicx]{revtex4-1}

\usepackage{amsmath}
\usepackage{bm}
\usepackage{graphicx}     

\usepackage [english]{babel}
\usepackage [autostyle, english = american]{csquotes}

\newcommand{\mt}[1]{\mathrm{#1}}

\newcommand{\ESO}{Mg$_{0.2}$Co$_{0.2}$Ni$_{0.2}$Cu$_{0.2}$Zn$_{0.2}$O}

\begin{document}

\preprint{}

\title{On the thermal and mechanical properties of \ESO~ across the high-entropy to entropy-stabilized transition} 



\author{Christina M. Rost}
\email[]{rostcm@jmu.edu}
\affiliation{Department of Physics and Astronomy, James Madison University, Harrisonburg, Virginia 22807, USA}

\author{Daniel L. Schmuckler}
\affiliation{Department of Physics and Astronomy, James Madison University, Harrisonburg, Virginia 22807, USA}

\author{Clifton Bumgardner}
\affiliation{Department of Mechanical and Aerospace Engineering, University of Virginia, Charlottesville, Virginia 22904, USA}

\author{Md Shafkat Bin Hoque}
\affiliation{Department of Mechanical and Aerospace Engineering, University of Virginia, Charlottesville, Virginia 22904, USA}

\author{David R. Diercks}
\affiliation{Department of Metallurgical and Materials Engineering, Colorado School of Mines, Golden, Colorado 80401, USA}

\author{John T. Gaskins}
\affiliation{Laser Thermal Incorporated, Charlottesville, VA 22902, USA}

\author{Jon-Paul Maria}
\affiliation{Department of Materials Science and Engineering, The Pennsylvania State University, University Park, Pennsylvania 16802, USA}

\author{Geoffrey L. Brennecka}
\affiliation{Department of Metallurgical and Materials Engineering, Colorado School of Mines, Golden, Colorado 80401, USA}

\author{Xiadong Li}
\affiliation{Department of Mechanical and Aerospace Engineering, University of Virginia, Charlottesville, Virginia 22904, USA}

\author{Patrick E. Hopkins}
\affiliation{Department of Materials Science and Engineering, University of Virginia, Charlottesville, Virginia 22904, USA}
\affiliation{Department of Physics, University of Virginia, Charlottesville, Virginia 22904, USA}
\affiliation{Department of Mechanical and Aerospace Engineering, University of Virginia, Charlottesville, Virginia 22904, USA}


\date{\today}

\begin{abstract}
As various property studies continue to emerge on high entropy and entropy-stabilized ceramics, we seek further understanding of property changes across the phase boundary between \enquote{high-entropy} and \enquote{entropy-stabilized}. The thermal and mechanical properties of bulk ceramic entropy stabilized oxide composition \ESO~are investigated across this critical transition temperature via the transient plane-source method, temperature-dependent X-ray diffraction, and nano-indentation. Thermal conductivity remains constant within uncertainty across the multi-to-single phase transition at a value of $\approx$~2.5 W/mK, while the linear coefficient of thermal expansion increases nearly 24 $\%$ from 10.8 to $\mt{14.1~x~10^{-6}} K^{-1}$. Mechanical softening is also observed across the transition.
\end{abstract}


\maketitle 

\section{Introduction}
The surge of scientific interest surrounding novel, multidimensional material compositions: medium and high entropy alloys (MEAs and HEAs) \cite{miracle_high-entropy_2017,murty_high-entropy_2019,yeh_high-entropy_2007,carroll_experiments_2015}, multicomponent alloys \cite{cantor_microstructural_2004}, multi-principal element alloys (MPEAs), complex concentrated alloys (CCAs), and their analogous ceramic-based compositions \cite{rost_entropy-stabilized_2015,harrington_phase_2019,gild_high-entropy_2016}, is building the pathways of understanding unique and complex structure-property relationships. Entropy stabilized oxides (ESOs)\cite{rost_entropy-stabilized_2015} are one such sub-class of high entropy oxides. ESOs add further complexity to the thermodynamic landscape of such materials, particularly those shown to be stabilized through entropic means \cite{mccormack_thermodynamics_2021,davies_thermodynamics_1981}\textemdash~ created through mixing equimolar amounts of several binary oxide components to overcome a positive enthalpy of mixing. The nature of this phase change is physically reversible. Above the critical temperature the system presents as a single-phase solid solution which, through quenching, remains metastable at room temperature. If slowly cooled below the critical transition, a multi-phase, high entropy composite is formed; where a secondary phase precipitates from the host matrix. Even with the presence of a secondary phase, the host still exhibits the key characteristic of a high entropy system in that the number of phases present is below predictions from the Gibbs phase rule \cite{gibbs_equilibrium_nodate}. Logically, similar to other systems exhibiting structural and/or compositional changes \cite{levin_phase_2001,de_la_calle_correlation_2008,mogni_oxygen_2013}, this phase precipitation and subsequent high-entropy restructuring would be expected to alter material properties.

Property studies continue to emerge on high entropy oxides with emphasis of functional applications such as catalysis and more \cite{toher_high-entropy_2022,sarkar_high_2018,albedwawi_high_2021}. To our knowledge, very little has been done with regards to property changes through the critical transition temperature. Here, we develop further understanding of and report on the thermal and mechanical property changes across the phase boundary between \enquote{high-entropy} and \enquote{entropy-stabilized} in bulk ceramic ESO \ESO.

\section{Materials and Methods}

ESO composition \ESO~(J14) was synthesized by combining equimolar amounts of MgO, NiO, CoO, CuO, and ZnO binary oxides, and mixing/milling using a SPEX SamplePrep (New Jersey, USA) high-energy ball mill with 5 mm yttria-stabilized zirconia milling media. Mixed powder was pressed into 12 mm diameter pellets by applying 5,000 lbs. of force using a Carver uniaxial hydraulic press (Indiana, USA). Green body pellets were reactively sintered at 1273 K for 2 hours in open atmosphere, followed by an air quench. X-ray diffraction measurements for phase analysis were performed using a PANalytical Empyrean (Almelo, Netherlands) diffractometer in Bragg-Brentano geometry with a Chi-Phi-XYZ stage and a PIX'cel detector.  

\textit{In situ} temperature measurements were done at the Analytical Instrumentation Facility (AIF) at North Carolina State University (Raleigh, NC) using an HTK 1200N oven heater stage for the Empyrean. Lattice parameters were determined using the GSAS-II analysis package \cite{toby_gsas-ii_2013}. A total of twenty-six X-ray powder diffraction patterns were collected. Each measurement spanned sixty degrees $2\theta$ with a step size of 0.0131 degrees. The sample was heated in 10 K increments at a rate of 10 K/min followed by a 10-minute equilibration time before measurement at each temperature. Two regions were measured: 983-1123 K for the multiphase and 1173-1273 K for the single-phase regimes, respectively. Shown in the temperature heating profile in Supplemental Figure S1, the sample was first heated through the single-phase transition, measured, then cooled to 973 K and held for 2 hours to ensure phase separation before re-measurement. The Rietveld Method \cite{rietveld_rietveld_2014} was used to find six constants, then a sequential fit was performed across eight variables. The final $ \mt{R_{wp}}$ values averaged between 2.68 and 4.33. A ten-coefficient Chebyschev background, sample displacement, micro-strain, initial lattice parameters, average cation occupancy, and thermal displacement values were refined at the minimum temperature of each regime. A sequential fit of lattice parameter changes was then performed on each set of diffractograms. For the multiphase data, refinement of phase fraction was also included. Results tables can be found in supporting information.

Nanoscale compositional and micro structural analysis through the phase transition was done using transmission electron microscopy (TEM). An FEI Helios Nanolab 600i was used to focus ion beam mill specimens using a lift-out method. TEM analysis was performed on an FEI Talos F200X instrument operated at 200 keV. 

Instrumented indentation, using the Oliver-Pharr methodology \cite{oliver_measurement_2004,pharr_improved_1992}, was used to extract the high temperature mechanical properties across the phase transition. Indentations were performed at temperatures up to 1223 K on a mechanically polished specimen on a nano indenter (MicroMaterials NanoTest Vantage) with a cubic boron nitride (cBN) Berkovich indenter tip. A series of ten indentations were performed on the specimen at each temperature as the sample was thermally cycled from ambient to 1223 K for three cycles within an open-air environment. A consistent heating and cooling rate of 1.6 K/min was used, and the specimen was allowed to stabilize at each temperature for one hour prior to performing the indentation.

The temperature-dependent thermal conductivity of a pair of J14 samples in both single and multi-phase were measured using the transient plane source system Hot Disk (TPS 3500, Thermtest) from 100 to 600 K. Additional details regarding the hot-disk method can be found elsewhere \cite{ridley_tailoring_2020,ding_thermal_2020}. Prior to measurements, both samples were annealed in air at 1273 K and 1073 K for one hour to promote stabilization in the single and multi-phase, respectively. The Hot Disk TPS 3500 was checked against two reference standards, stainless-steel and a BK7 window glass, and the measured values were found to be in good agreement with literature \cite{ho_electrical_nodate, assael_thermal_2005,braun_steady-state_2019,yuan_thermal_2011}. The J14 samples were $\approx$~85 \% dense, estimated via mass/volume calculations, and thermal conductivity values were subsequently adjusted using the Maxwell-Garnett model shown in Equation~\ref{k_eqn}\cite{nan_effective_1997, ridley_tailoring_2020} below, where $\phi$ is the relative porosity percentage of the samples. 

\begin{equation}
\displaystyle 
k_{solid}= \frac{k_{porous}}{\frac{1-\phi}{1+\phi}}
\label{k_eqn}
\end{equation}

\section{Results and Discussion}

\begin{figure}[h]%
	\includegraphics*[scale=.5]{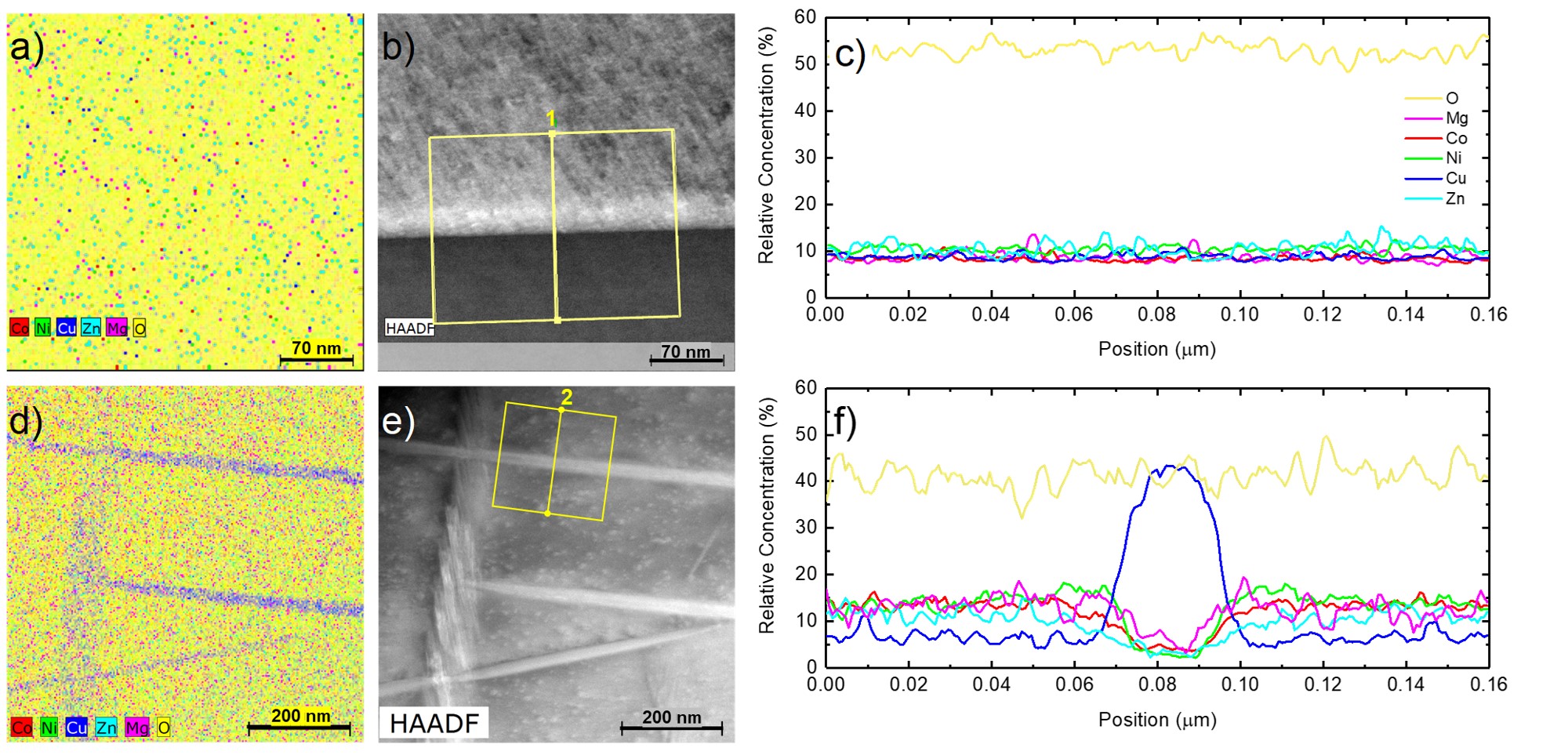}
	\centering
	\caption{TEM micrographs and corresponding EDX-measured elemental distributions of both single phase rocksalt and phase-separated J14. Panel a shows the EDX elemental distribution map, b is the TEM image associated with the EDX mapping, and c is an EDX line scan confirming composition homogeneity across a grain boundary for the single phase; characteristic of the expected rocksalt solid solution. Similarly, panels d-f show the elemental distribution, TEM image, and line scan across the phase boundary, respectively, for the multi-phase regime where needle-like grains consisting primarily of CuO precipitate from the rocksalt host.}
	\label{fig:EDS}
\end{figure}

Figure ~\ref{fig:EDS} shows the microstructure and chemical distribution of J14 above and below the transition temperature, characterized via transmission electron microscopy (TEM) and energy dispersive spectroscopy (EDX). Consistent with previous results \cite{hong_microstructural_2019}, the single-phase rocksalt solution transitions to a two phase system, where CuO needles precipitate from the rocksalt host. 

\begin{figure}[h]%
	\centering
	\includegraphics*[scale=.3]{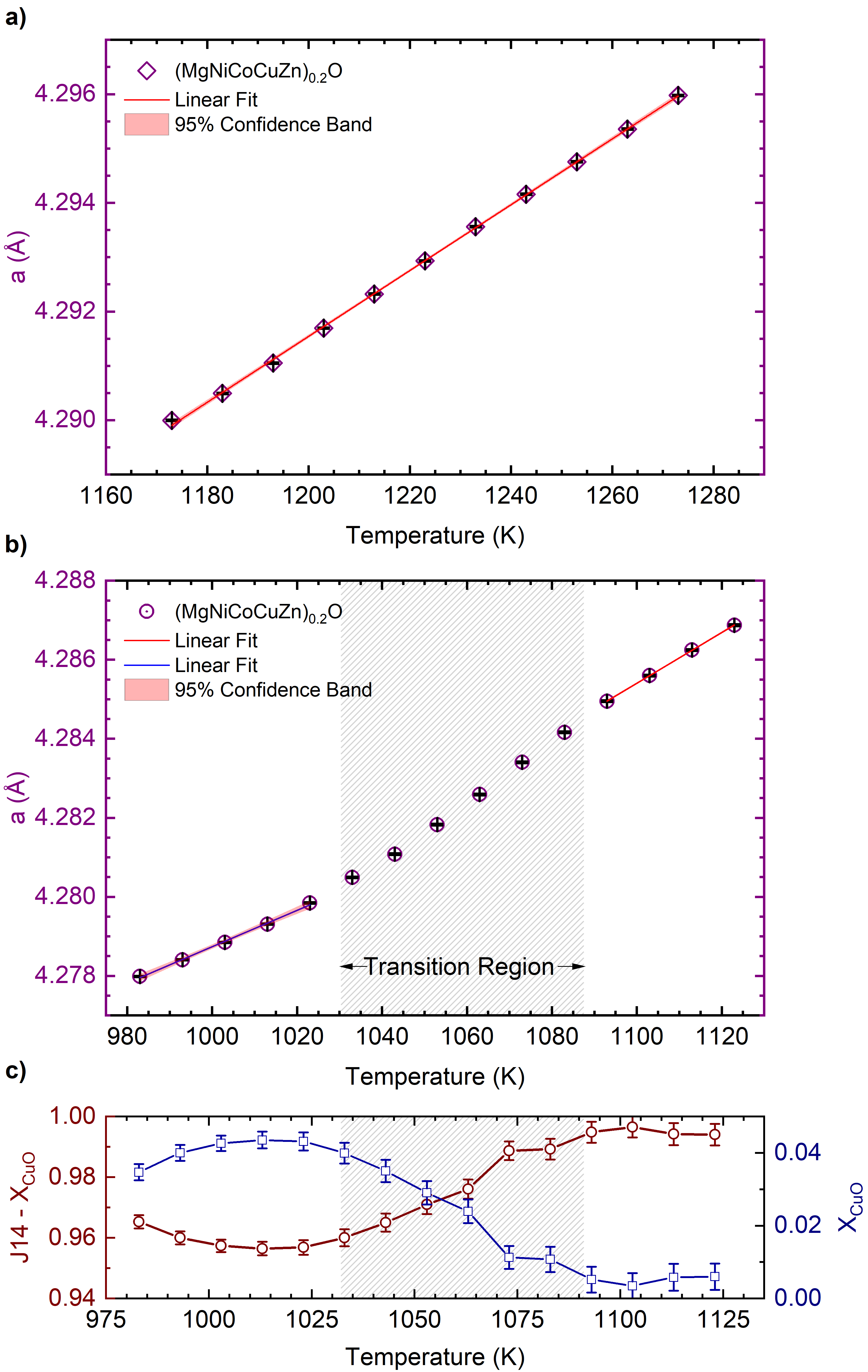}
	\caption{Panels a and b plot lattice parameter versus temperature below and above the multi-to-single phase transition temperature for \ESO, respectively. Included with these plots are the linear fits and associated 95 \% confidence bands used for determining the CTE in each region of interest. Panel c plots the change in phase fraction across the transition temperature as determined though refinement, suggesting that approximately 3-4 \% of CuO precipitates from J14 below the transition. Between 1030 K and 1090 K, CuO re-incorporates into the host matrix.
	}
	\label{fig:CTE_2}
\end{figure}

Due to the isotropic nature of the J14 phase, we are able to reduce thermal expansion to one dimension. The linear coefficient of thermal expansion (CTE) for J14 shown in Figure ~\ref{fig:CTE_2}, plots lattice parameter versus temperature for a) single phase J14 and b) Cu deficient J14, as found in the two-phase regime. The slope is used to determine the CTE based on the basic equation:

\begin{equation}
\displaystyle 
\alpha_L= \frac{1}{L}\frac{\Delta L}{\Delta T}
\label{CTE}
\end{equation}

where $\alpha_L$ is the linear CTE, L is the lower limit of the lattice parameter in the measurement, and $\Delta L/\Delta T$ is the slope. Based on these data, we find the CTE for single phase J14 (Figure ~\ref{fig:CTE_2}a) to be $\mt{ 14.12~ \pm~0.09~x~10^{-6}} ~K^{-1}$. The two-phase temperature regime (Figure ~\ref{fig:CTE_2}b) is somewhat more complicated: Cu is actively dissolving back into the rocksalt host across the transition back to single phase concurrently with its thermal expansion, creating a superimposed function on the CTE as the host makes room for additional cations. Rietveld refinement of the weight fractions of CuO and J14 as a function of temperature, shown in Figure ~\ref{fig:CTE_2}c suggests a transition range from approximately 1030 K to 1085 K, where the remaining CuO nearly completely dissolves. This temperature range agrees well with previous observations of the transition from multi-phase to single phase J14 \cite{rost_entropy-stabilized_2015}. It can be seen that above this transition region, the coefficient of thermal expansion is found to be $\mt{ 14.97~\pm~0.09~x~10^{-6}} ~K^{-1}$, an approximately 4.7 $\%$ difference from the single phase regime. Referring back to Figure ~\ref{fig:CTE_2}c, this could be due to 0.6 weight $\%$ Cu still slowly diffusing into the host lattice.  Below the transition range, a larger concentration of CuO ($\approx$4 weight $\%$)  exists outside of J14, and the CTE of the primary phase drops to  $\mt{ 10.8~\pm~0.3~x~10^{-6}} ~K^{-1}$.

\begin{figure}[]
	\centering
	\includegraphics*[scale=.3]{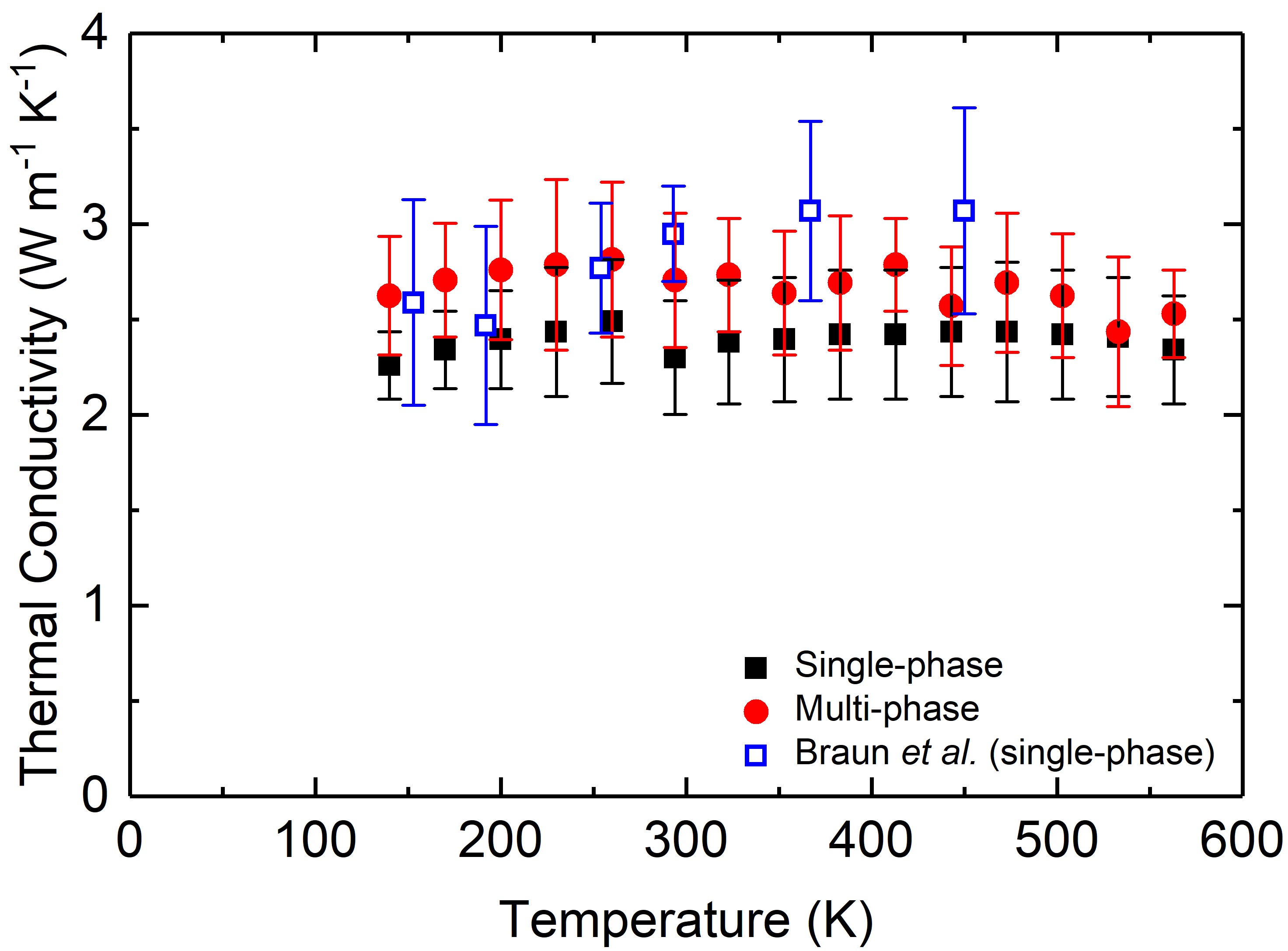}
	\caption{\label{fig:ESO_k}Temperature dependent thermal conductivity of J14 above and below the critical temperature appears to be constant within uncertainty regardless of phase. These values agree with those reported previously, approximately $\mt{ 2.5 - 3} ~W m^{-1} K^{-1}$. \cite{braun_charge-induced_2018}}%
\end{figure}

Figure~\ref{fig:ESO_k} shows the thermal conductivity of the single-phase and multi-phase J14 samples as a function of temperature. These thermal conductivity values were measured using the transient plane source method \cite{gustavsson_thermal_1994} via a hot-disk. 

Within uncertainty, the thermal conductivity of the single-phase and multi-phase samples are nearly the same and span $\mt{ 2.5 - 3} ~W m^{-1} K^{-1}$, which is in agreement with the $\mt{ 2.95~\pm~0.25} ~W m^{-1} K^{-1}$  reported by Braun, et al. \cite{braun_charge-induced_2018}  . The primary heat carrier in both phases of J14 is the phonons. The near same thermal conductivity observed here between the two phases indicate that the precipitation of CuO in the multi-phase sample does not increase phonon scattering significantly. Additionally, both phases exhibit an  amorphous-like thermal conductivity trend as the temperature increases from 100 to 600 K. Such amorphous-like trends is expected in ESOs \cite{braun_charge-induced_2018}. The presence of five cations in the crystal lattice causes a significant amount of phonon scattering \cite{yan_hf0_2018}. Further, the overall porosity of the samples provides additional sites for phonon scattering \cite{chen_high_2019}. Together, these two mechanisms lead to the amorphous-like thermal conductivity in single-phase and multi-phase ESOs.

\begin{figure}[t]
	\includegraphics[scale = .3]{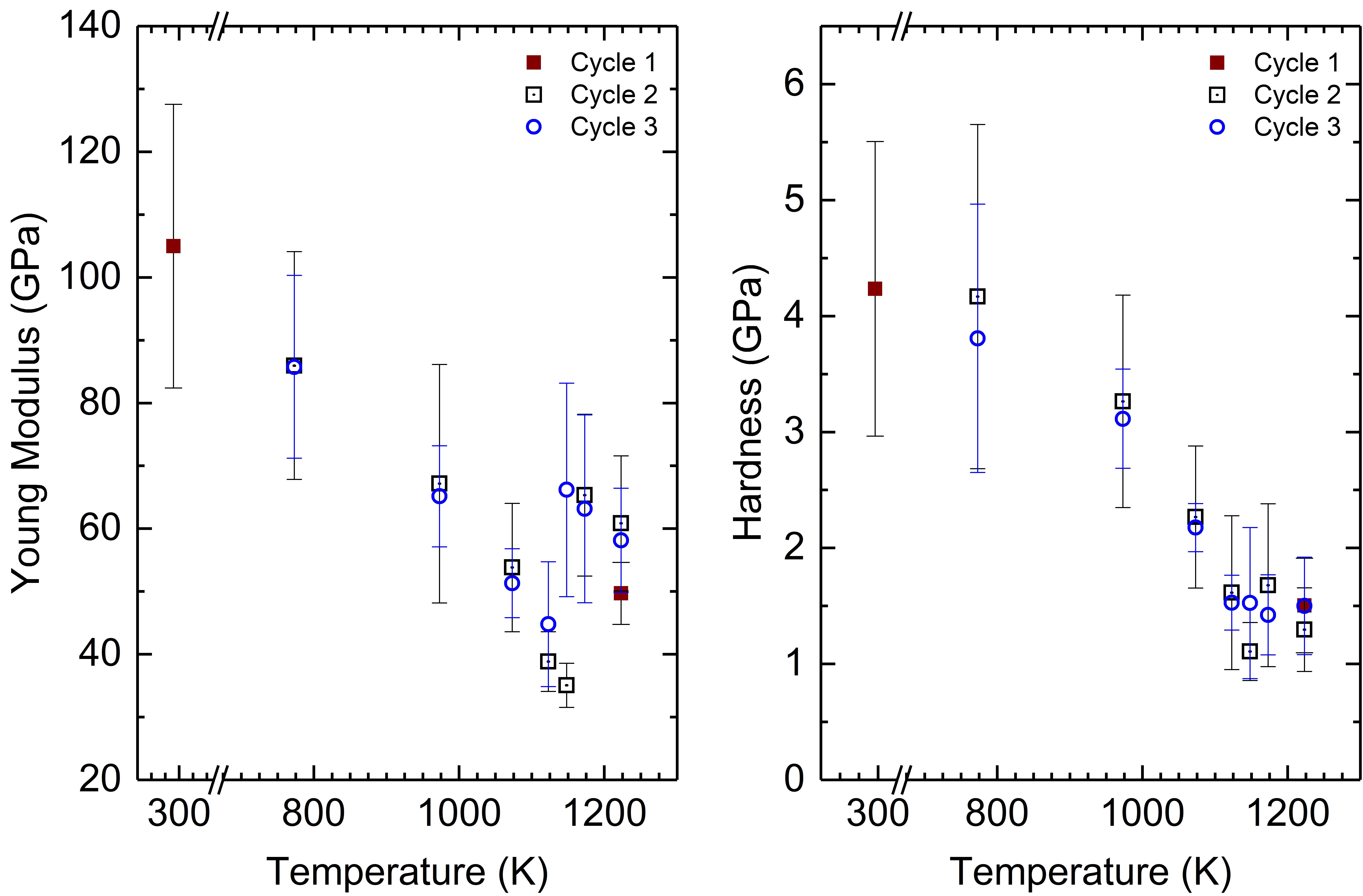}
	\centering
	\caption{Elastic modulus measurements derived from indentation. A distinctive ‘step’ feature observed at about 1150 K, indicative of a temperature-activated stiffening mechanism. Hardness measurements derived from indentation, showing significant softening of the material above 775 K.}
	\label{fig:EM}
\end{figure}

As shown in Figure ~\ref{fig:EM}, the specimen was observed to retain its mechanical properties through 775 K, but between 775 and 1125 K, the specimen experienced significant softening before stabilizing above 1125 K. Similarly, the elastic modulus in Figure ~\ref{fig:EM} decreased significantly within this same interval, 875 K to 1125 K; however, there was a distinct ‘step’ feature occurring around 1125 K at which point the modulus increased about 25~\%. This distinctive step was not present in the hardness measurement, only within the modulus measurements, indicative of a lattice-stiffening, temperature-driven mechanism. Furthermore, this step was fully detectable in both cycle 2 and 3, demonstrating the underlying mechanism to be reversible. These indentation measurements indicate temperature-driven activation of a reversible phase transformation at about 1125 K, after which key mechanical properties may be stabilized, even slightly recovered. Hardness and modulus measurements are frequent indicators of toughening and stiffening mechanisms respectively and have been used in literature to demonstrate active phase transformations in metals and ceramics\cite{wang_first-principles_2011,orlovskaya_mechanical_2000,cheng_mechanical_1996}.

\section{Conclusions}

The resulting effects on thermal and mechanical properties from the reversible transformations observed in entropy-stabilized oxides were investigated. Indentation results support previously noted phase transformation reversibility within these entropy-stabilized oxides, while temperature-dependent XRD results confirm this mechanism is temperature-driven, not primarily stress or indentation-load driven.  Importantly, the stability and even limited recovery of mechanical properties after this structure-property transformation, combined with the constant thermal conductivity, suggest this effect may be reliably exploited for mid-to-high temperature application of entropy-stabilized oxides. The substantial difference in linear CTE within regions surrounding the transition temperature may cause difficulty in applying J14 as a coating, however to fully understand such applied behavior more study would be required.

\section*{Acknowledgements\\}
C.M.R. and D.L.S. would like to acknowledge funding by 4-VA, a collaborative partnership for advancing the Commonwealth of Virginia. C.M.R. also gratefully acknowledges partial support from NSF through the Materials Research Science and Engineering Center DMR-2011839. Md.S.B.H. and P.E.H. are appreciative for support from the Office of Naval Research, Grant Number N00014-21-1-2477. This work was performed in part at the Analytical Instrumentation Facility (AIF) at North Carolina State University, which is supported by the State of North Carolina and the National Science Foundation (award number ECCS-1542015). The AIF is a member of the North Carolina Research Triangle Nanotechnology Network (RTNN), a site in the National Nanotechnology Coordinated Infrastructure (NNCI).

\section*{Conflict of Interest\\}

The authors declare no conflict of interest.

\section*{Supporting Information\\}

Supporting information tables are available.



%
%

%


\bibliography{CTE_Mech_bibliography_biblatex_v3}

\end{document}